\documentclass[a4paper]{article}
\RequirePackage[english]{babel}
\RequirePackage[latin1]{inputenc}
\RequirePackage[T1]{fontenc}
\RequirePackage{mathrsfs}
\RequirePackage{amsmath}
\RequirePackage{amssymb}
\RequirePackage{amsbsy}
\RequirePackage{bm}
\begin{document}
\title{\bf{Metric Solutions in Torsionless Gauge\\ for Vacuum Conformal Gravity}}
\author{Luca Fabbri}
\date{}
\maketitle
\begin{abstract}
In a recent paper we have established the form of the metric-torsional conformal gravitational field equations, and in the present paper we study their vacuum configurations; we will consider a specific situation that will enable us to look for the torsionless limit: two types of special exact solutions are found eventually. A discussion on general remarks will follow.
\end{abstract}
\section*{Introduction}
In defining a relativistic theory of gravity in which to implement conformal transformations accounting for length scaling, the main goal is to find a possible model for gravity that is also renormalizable, so to eventually be able to treat the problem of gravitational quantization, as it has been discussed by Stelle in reference \cite{s}: according to this procedure, one has to find solutions in the form of gravitational waves expanded in terms of monochromatic plane waves on which to transfer the commutation relations set by the rules of quantization, and in the past solutions in the form of gravitational waves have been discussed extensively, mainly by Mannheim, as for instance in \cite{m}, and also by Paranjape and collaborators, as for example in \cite{b-p} and \cite{f-p}. On the other hand however, if one wants to consider gravity in its most exhaustive coupling to matter, then one must consider a background in which not only curvature but also torsional degrees of freedom are considered, so that not only the energy but also the spin is coupled to the background: in such frameworks, curvature and torsion can be seen as the strengths of the potentials arising after gauging the Poincar\'{e} group of the spacetime roto-translations \cite{Hehl:1994ue,h-o}, so that we may interpret curvature and torsion as what gives rise for the spacetime continuum to disclinations and dislocations \cite{k}. Thus, the problem of having a theory of gravitation suitable of being quantized in its most complete form relies upon the problem of having a theory of gravitation that is renormalizable and in which both energy and spin density are present, and hence a theory of gravitation in which conformal invariance is implemented for both curvature and torsion, as discussed in \cite{sh}.

The problem of such an approach was that there was no known way to obtain for the Weyl conformal curvature in the purely metric case any torsional extension, and thus no way in which the Weyl conformal gravity could be obtained as a torsionless limit: nevertheless, such an extension has been found in recent papers \cite{f,f/1,f/2}. Then, the metric-torsional conformal gravitational field equations have been given, and their consistency has been checked in terms of the conservation laws that have to be satisfied by the spin and energy conformal quantities of any matter fields, once the conformally invariant matter field equations are given. The next step is then to look for special situations like those we have in absence of any matter field for the vacuum configurations.

It is well-known that Schwarzschild-like vacuum solutions for Weyl conformal gravity are in fact extensions of the Schwarzschild vacuum solution of the Einsteinian gravitation \cite{m-k}; on the other hand, it is essential for the present torsional extension of the metric Weyl gravity to find vacuum solutions in order to see how they may further extend the purely metric Weyl gravity, and so the Einsteinian gravitation, or if instead they are a more restricted class of solution with respect to Weyl gravity. In this paper, we shall consider such solutions and their relationships in two cases of special symmetries.
\section{The Metric-Torsional Conformal Gravitation}
In this paper, the Riemann-Cartan metric-torsional geometry is defined in terms of a metric $g_{\alpha\beta}$ and a connection $\Gamma^{\mu}_{\alpha\sigma}$ that is metric-compatible: the condition of metric-compatibility means that by applying to the metric tensor the covariant derivative associated to the connection the result vanishes; the connection is not symmetric in the two lower indices and its antisymmetric part in those two indices is a tensor known as Cartan torsion tensor
\begin{eqnarray}
&Q^{\sigma}_{\phantom{\sigma}\rho\alpha}=\Gamma^{\sigma}_{\rho\alpha}-\Gamma^{\sigma}_{\alpha\rho}
\label{Cartan}
\end{eqnarray}
so that the connection is decomposable in a unique way according to the
\begin{eqnarray}
&\Gamma^{\sigma}_{\phantom{\sigma}\rho\alpha}=
\frac{1}{2}g^{\sigma\theta}[Q_{\rho\alpha\theta}+Q_{\alpha\rho\theta}+Q_{\theta\rho\alpha}
+(\partial_{\rho}g_{\alpha\theta}+\partial_{\alpha}g_{\rho\theta}-\partial_{\theta}g_{\rho\alpha})]
\label{connection}
\end{eqnarray}
showing that because of Cartan torsion the metric and the metric-compatible connections are indeed independent. An equivalent formalism can be introduced, in which we consider the constant Minkowskian metric $\eta_{ij}$ and a basis of vierbein $e_{\alpha}^{i}$ such that we have the relationship $e_{\alpha}^{p}e_{\nu}^{i}\eta_{pi}=g_{\alpha\nu}$ together with the spin-connection $\omega^{ip}_{\phantom{ip}\alpha}$ and where vierbein and spin-connection are again taken to be independent: the correspondent metric-compatibilities spell that by applying to the vierbein and Minkowskian metric the covariant derivative associated to the spin-connection both yield zero; for such a spin-connection it is not possible to define an analogous of torsion although we have the relationship
\begin{eqnarray}
&Q^{i}_{\phantom{i}\alpha\rho}
=-\left(\partial_{\alpha}e_{\rho}^{i}-\partial_{\rho}e_{\alpha}^{i}
+e_{\rho}^{p}\omega^{i}_{\phantom{i}p\alpha}-e_{\alpha}^{p}\omega^{i}_{\phantom{i}p\rho}\right)
\label{torsion}
\end{eqnarray}
in terms of vierbein and a spin-connection given according to
\begin{eqnarray}
&\omega^{i}_{\phantom{i}p\alpha}
=e^{i}_{\sigma}(\Gamma^{\sigma}_{\rho\alpha}e^{\rho}_{p}+\partial_{\alpha}e^{\sigma}_{p})
\label{spin-connection}
\end{eqnarray} 
with a spin-connection that is antisymmetric $\omega^{ip}_{\phantom{ip}\alpha}\!
=\!-\omega^{pi}_{\phantom{pi}\alpha}$ and showing that vierbein and spin-connection are in fact independent. The former formalism with Latin letters and the latter formalism with Greek letters are called spacetime and world formalism, and they are equivalent; in these the independence between metric and connection is equivalent to the independence between vierbein and spin-connection. For a more extensive introduction we refer to \cite{f}.

The conformal transformation is given by requiring that the line element is stretched by a given function $\sigma$ and therefore we have that for the metric it is expressed as usual by $g_{\alpha\beta}\rightarrow\sigma^{2}g_{\alpha\beta}$ while by defining $\ln{\sigma}=\phi$ we have that for the torsion tensor it is given by the following transformation law
\begin{eqnarray}
&Q^{\sigma}_{\phantom{\sigma}\rho\alpha}\rightarrow Q^{\sigma}_{\phantom{\sigma}\rho\alpha}
+q(\delta^{\sigma}_{\rho}\partial_{\alpha}\phi-\delta^{\sigma}_{\alpha}\partial_{\rho}\phi)
\end{eqnarray}
in terms of the parameter $q$ called conformal charge, and we have that its only contraction  $Q^{\rho}_{\phantom{\rho}\rho\nu}=Q_{\nu}$ is the only irreducible part that inherits the conformal transformation; from the relationship (\ref{connection}) it is possible to see what is the conformal transformation for the connection in general. Given that there is no conformal transformation for the constant Minkowskian matrix then the conformal transformation for the vierbein is $e_{\alpha}^{k}\rightarrow\sigma e_{\alpha}^{k}$ as it is clear; by employing the relationship (\ref{spin-connection}) it is possible to get the conformal transformation for the spin-connection in general. Here we will only focus on the case in which the conformal charge is in general not vanishing, as discussed in \cite{f} the simplest case in which the conformal charge vanishes cannot be treated as a simple limit of this model and therefore a parallel study must be done in a different work.

In this framework, the Riemann metric-torsional curvature tensor is
\begin{eqnarray}
\nonumber
&G^{i}_{\phantom{i}k\mu\nu}=G^{\rho}_{\phantom{\rho}\xi\mu\nu}e^{i}_{\rho}e^{\xi}_{k}=\\
\nonumber
&=(\partial_{\mu}\Gamma^{\rho}_{\xi\nu}-\partial_{\nu}\Gamma^{\rho}_{\xi\mu}
+\Gamma^{\rho}_{\sigma\mu}\Gamma^{\sigma}_{\xi\nu}
-\Gamma^{\rho}_{\sigma\nu}\Gamma^{\sigma}_{\xi\mu})e^{i}_{\rho}e^{\xi}_{k}\equiv\\
&\equiv\partial_{\mu}\omega^{i}_{\phantom{i}k\nu}-\partial_{\nu}\omega^{i}_{\phantom{i}k\mu}
+\omega^{i}_{\phantom{i}a\mu}\omega^{a}_{\phantom{a}k\nu}
-\omega^{i}_{\phantom{i}a\nu}\omega^{a}_{\phantom{a}k\mu}
\label{Riemanncurvature}
\end{eqnarray}
with the implicit presence of Cartan torsion tensor; it is antisymmetric in both the first and second couple of indices, and accordingly it has one independent contraction $G^{\rho}_{\phantom{\rho}\mu\rho\nu}\!= \!G^{i}_{\phantom{i}\mu\rho\nu}e^{\rho}_{i}\!=\!G_{\mu\nu}$ with $G_{\eta\nu}g^{\eta\nu}\!=\!G$ setting our convention.

From the conformal transformation of Cartan torsion tensor and the definition of Riemann curvature tensor it is possible to see what the conformal transformation of the Riemann curvature tensor must be, and it is easy to see that because of the presence of covariant derivatives of torsion additional terms are present which render the conformal transformation of the Riemann curvature tensor more complicated than the usual one, in fact so much more complicated that its irreducible part fails to be conformally covariant; this is due to the fact that within the Riemann curvature tensor, Cartan torsion tensor enters implicitly through the connection, but it is also possible to add the Cartan torsion tensor explicitly so to get the expression of the modified metric-torsional tensor
\begin{eqnarray}
&M_{\alpha\beta\mu\nu}
=G_{\alpha\beta\mu\nu}+(\frac{1-q}{3q})(Q_{\beta}Q_{\alpha\mu\nu}-Q_{\alpha}Q_{\beta\mu\nu})
\label{curvature}
\end{eqnarray}
with the same symmetries of Riemann curvature tensor but also such that
\begin{eqnarray}
&T_{\alpha\beta\mu\nu}=M_{\alpha\beta\mu\nu}
-\frac{1}{2}(M_{\alpha[\mu}g_{\nu]\beta}-M_{\beta[\mu}g_{\nu]\alpha})
+\frac{1}{12}M(g_{\alpha[\mu}g_{\nu]\beta}-g_{\beta[\mu}g_{\nu]\alpha})
\label{conformalcurvature}
\end{eqnarray}
not only has the same symmetries and it is irreducible but it is also conformally covariant as a direct computation would easily show: given this tensor, we notice that in building the coordinate-conformal invariants we have three possible contractions, so that in terms of three parameters $A$, $B$, $C$ the most general invariant is given by $AT^{\alpha\beta\mu\nu}T_{\alpha\beta\mu\nu}+ BT^{\alpha\beta\mu\nu}T_{\mu\nu\alpha\beta}+ CT^{\alpha\beta\mu\nu}T_{\alpha\mu\beta\nu}$ and thus it is useful to define the parametric tensor given by the following expression
\begin{eqnarray}
&P_{\alpha\beta\mu\nu}=AT_{\alpha\beta\mu\nu}+BT_{\mu\nu\alpha\beta}+\frac{C}{4}(T_{\alpha\mu\beta\nu}-T_{\beta\mu\alpha\nu}+T_{\beta\nu\alpha\mu}-T_{\alpha\nu\beta\mu})
\label{parametricconformalcurvature}
\end{eqnarray}
in terms of the parameters $A$, $B$, $C$, still with the same symmetries and irreducible and also conformally covariant in $(1\!+\!3)$-dimensional spacetimes and such that the quantity $T^{\alpha\beta\mu\nu}P_{\alpha\beta\mu\nu}$ is the most general invariant we may write.

Before proceeding to write the gravitational action, it may be instructive to see that for a generic gravitational Lagrangian the above equivalent formalisms will yield equivalent systems of field equations: to see that, we notice that the variation of a gravitational Lagrangian $L$ is obtained by the variation either with respect to the metric and torsion according to the following expression
\begin{eqnarray}
\frac{\delta L}{\delta g_{\mu\nu}}=E^{\mu\nu}\\
\frac{\delta L}{\delta Q^{\alpha}_{\phantom{\alpha}\mu\nu}}=K_{\alpha}^{\phantom{\alpha}\mu\nu}
\end{eqnarray}
or with respect to vierbein and spin-connection according to the forms
\begin{eqnarray}
\frac{\delta L}{\delta e_{\mu}^{j}}=B^{\mu}_{\phantom{\mu}j}\\
\frac{\delta L}{\delta \omega^{ij}_{\phantom{ij}\nu}}=\Sigma^{\nu}_{\phantom{\nu}ij}
\end{eqnarray}
and so by means of the identities $\delta g_{\mu\nu}\!=\!(\eta_{ab}e^{b}_{\phantom{b}\nu} \delta e^{a}_{\phantom{a}\mu}\!+\!\eta_{ab}e^{b}_{\phantom{b}\mu}\delta e^{a}_{\phantom{a}\nu})$ together with the identities $\delta  Q^{\theta}_{\phantom{\theta}\mu\rho} \!=\!-\delta^{\beta}_{[\mu}e^{k}_{\rho]} e^{\theta}_{i}\eta_{kj}\delta \omega^{ij}_{\phantom{ij}\beta}$ and $\delta Q^{\theta}_{\phantom{\theta}\mu\rho}\!=\!-e^{\theta}_{i}D_{[\mu}\delta e^{i}_{\rho]}$ we have that
\begin{eqnarray}
&\Sigma^{\nu}_{\phantom{\nu}ij}\!=\!K_{ij}^{\phantom{ij}\nu}-K_{ji}^{\phantom{ji}\nu}
\label{structural}\\
&\frac{1}{2}B^{\rho}_{\phantom{\rho}a}\!=\!E^{\rho}_{\phantom{\rho}a}
+Q_{\mu}K_{a}^{\phantom{a}\mu\rho}+D_{\mu}K_{a}^{\phantom{a}\mu\rho}
\label{local}
\end{eqnarray}
and the inverted
\begin{eqnarray}
&K_{kij}\!=\!\frac{1}{2}(\Sigma_{ijk}\!-\!\Sigma_{jik}\!-\!\Sigma_{kij})
\label{structureate}\\
&2E^{\rho\alpha}\!=\!B^{\rho\alpha}
\!+\!Q_{\mu}(\Sigma^{\rho\mu\alpha}\!-\!\Sigma^{\mu\rho\alpha}\!-\!\Sigma^{\alpha\rho\mu})
\!+\!D_{\mu}(\Sigma^{\rho\mu\alpha}\!-\!\Sigma^{\mu\rho\alpha}\!-\!\Sigma^{\alpha\rho\mu})
\label{localized}
\end{eqnarray}
showing that the two stationary minima $K_{\alpha\beta\mu}\!=\!0$ and $E_{\rho\nu}\!=\!0$ occur if and only if the stationary minima $\Sigma^{\nu}_{\phantom{\nu}ij}\!=\!0$ and $B^{\rho}_{\phantom{\rho}a}\!=\!0$ occur, demonstrating the equivalence of the two systems of field equations, as expected; notice however that although the field equations $K_{\alpha\beta\mu}\!=\!0$ and $\Sigma^{\nu}_{\phantom{\nu}ij}\!=\!0$ are identical up to an index rearrangement, field equations $E_{\rho\nu}\!=\!0$ and $B^{\rho}_{\phantom{\rho}a}\!=\!0$ are different because of the presence of the divergence of the torsion-spin field equation, as it would have to be expected since $E_{\rho\nu}\!=\!0$ are the symmetric part of $B^{\rho}_{\phantom{\rho}a}\!=\!0$ as it has already been discussed in the literature, and for which the reader interested in further details may have a look for instance at reference \cite{Hehl:1994ue}. It is also important to notice that although this holds in general, in what follows we will provide an explicit proof of this fact in the specific case of conformal gravity we will study.

Now we may proceed to choose what we have anticipated to be the gravitational conformal action given in the most general case by the expression
\begin{eqnarray}
&S=\int[kT^{\alpha\beta\mu\nu}P_{\alpha\beta\mu\nu}+L_{\mathrm{matter}}]\sqrt{|g|}dV
\label{action}
\end{eqnarray}
with constant $k$ complemented by the material Lagrangian and where it is over the volume of the spacetime that the integral is taken; we have shown that any variational procedure yields equivalent systems of field equations, and so we may choose whichever we desire and in this paper we will perform the variation with respect to the vierbein and the spin-connection obtaining the field equations
\begin{eqnarray}
\nonumber
&2k[P^{\theta\sigma\rho\alpha}T_{\theta\sigma\rho}^{\phantom{\theta\sigma\rho}\mu}
-\frac{1}{4}g^{\alpha\mu}P^{\theta\sigma\rho\beta}T_{\theta\sigma\rho\beta}
+P^{\mu\sigma\alpha\rho}M_{\sigma\rho}+\\
\nonumber
&+(\frac{1-q}{3q})
(D_{\nu}(2P^{\mu\rho\alpha\nu}Q_{\rho}
-g^{\mu\alpha}P^{\nu\theta\rho\sigma}Q_{\theta\rho\sigma}
+g^{\mu\nu}P^{\alpha\theta\rho\sigma}Q_{\theta\rho\sigma})+\\
&+Q_{\nu}(2P^{\mu\rho\alpha\nu}Q_{\rho}
-g^{\mu\alpha}P^{\nu\theta\rho\sigma}Q_{\theta\rho\sigma}
-P^{\mu\nu\rho\sigma}Q^{\alpha}_{\phantom{\alpha}\rho\sigma}))]=\frac{1}{2}T^{\alpha\mu}
\label{energy}\\
\nonumber
&4k[D_{\rho}P^{\alpha\beta\mu\rho}+Q_{\rho}P^{\alpha\beta\mu\rho}
-\frac{1}{2}Q^{\mu}_{\phantom{\mu}\rho\theta}P^{\alpha\beta\rho\theta}-\\
&-(\frac{1-q}{3q})(Q_{\rho}P^{\rho[\alpha\beta]\mu}
-\frac{1}{2}Q_{\sigma\rho\theta}g^{\mu[\alpha}P^{\beta]\sigma\rho\theta})]=S^{\mu\alpha\beta}
\label{spin}
\end{eqnarray}
in terms of the parameter $q$ and the constant $k$ and where $T^{\mu\nu}$ and $S^{\rho\mu\nu}$ are the energy and spin densities of the conformal matter field: finally by taking into account the Jacobi-Bianchi identities, the field equations (\ref{energy}-\ref{spin}) are converted into the usual conservation laws given according to the expressions
\begin{eqnarray}
&D_{\mu}T^{\mu\rho}+Q_{\mu}T^{\mu\rho}-T_{\mu\sigma}Q^{\sigma\mu\rho}
+S_{\beta\mu\sigma}G^{\sigma\mu\beta\rho}=0
\label{conservationlawenergy}\\
&D_{\rho}S^{\rho\mu\nu}+Q_{\rho}S^{\rho\mu\nu}
+\frac{1}{2}T^{[\mu\nu]}=0
\label{conservationlawspin}
\end{eqnarray}
with trace condition as another conservation law
\begin{eqnarray}
&(1-q)(D_{\mu}S_{\nu}^{\phantom{\nu}\nu\mu}+Q_{\mu}S_{\nu}^{\phantom{\nu}\nu\mu})
+\frac{1}{2}T_{\mu}^{\phantom{\mu}\mu}=0
\label{trace}
\end{eqnarray}
satisfied once conformal matter field equations are given, and where the fact that the general conservation laws (\ref{conservationlawenergy}-\ref{conservationlawspin}) are now accompanied by an additional conservation law for the trace (\ref{trace}) comes from the fact that now there is not only general coordinate and special Lorentz invariance but also conformal invariance for the gravitational action we want to study. Also remark that if we do not neglect the torsion and its strong conformal transformations then for non-traceless spin density tensors and unless we are in the particular case where the conformal charge is one we have that the conservation law for the trace is not a mere constraint on the energy density but a truly dynamical conservation law, analogous to the one linking the spin density to the antisymmetric part of the energy density in a dynamical way, and we regard this feature as an improvement with respect to the torsionless conformally invariant theory.
\section{Metric Solutions in Torsionless Gauge\\ for the Vacuum in Conformal Gravitation}
This is the gravitational model presented in \cite{f,f/1,f/2}, and in what follows we are going to apply it to the situation of vacuum configurations, in which the matter field is absent so that both energy and spin density tensors can be set to zero, and the vacuum field equations will turn out to be more manageable.

In these vacuum configurations, a first problem we may now try to solve is whether the case in which torsion vanishes would reduce to the case in which torsion was never present as in the purely metric approach; of course, because the conformal transformation for torsion $Q^{\mu}_{\phantom{\mu}\alpha\sigma}$ is entirely loaded on the trace vector decomposition $Q_{\nu}$, then the axial vector and the remaining decompositions are conformally invariant: this means that it is meaningless to set the the trace vector $Q_{\nu}$ to zero because it can always be produced by a conformal transformation, and therefore the other two decompositions are the only ones of which we can meaningfully require the vanishing. Thus said, in the following we shall study the case that is the closest to the one in which torsion vanishes, that is when torsion is entirely given by its trace vector part as
\begin{eqnarray}
&Q^{\mu}_{\phantom{\mu}\alpha\sigma}
\equiv\frac{1}{3}(\delta^{\mu}_{\alpha}Q_{\sigma}-\delta^{\mu}_{\sigma}Q_{\alpha})
\end{eqnarray}
that is the one containing all conformal transformation properties, setting to zero all other parts, which are conformally invariant. With this requirement, it is straightforward to see that $T_{\rho\sigma\mu\nu}\!\equiv\!C_{\rho\sigma\mu\nu}$ that is the conformal curvature defined here identically reduces to the conformal curvature of the purely metric case given by the Weyl curvature defined as usual; so the metric and torsional degrees of freedom decouple at least within the conformal curvature.

Also we have that further simplifications come by choosing special fine-tunings, in this case for the parameters $A$, $B$, $C$; however, because the conformal curvature is now the Weyl curvature, with all its symmetry properties of indices transposition, then there is no loss of generality in choosing the parameters in such a way that $P_{\rho\sigma\mu\nu}\!\equiv\!T_{\rho\sigma\mu\nu}\!
=\!C_{\rho\sigma\mu\nu}$ so to have the parametric conformal curvature reduced to the conformal Weyl curvature. With these requirements, the conformal field equations (\ref{energy}-\ref{spin}) respectively reduce to the form
\begin{eqnarray}
\nonumber
&C^{\theta\sigma\rho\alpha}C_{\theta\sigma\rho}^{\phantom{\theta\sigma\rho}\mu}
-\frac{1}{4}g^{\alpha\mu}C^{\theta\sigma\rho\beta}C_{\theta\sigma\rho\beta}
+C^{\mu\sigma\alpha\rho}R_{\sigma\rho}+\\
&+\frac{2}{9q^{2}}C^{\mu\sigma\alpha\rho}Q_{\rho}Q_{\sigma}
+\frac{2}{3q}C^{\mu\sigma\alpha\rho}\nabla_{\rho}Q_{\sigma}=0
\label{energyvacuum}\\
&\nabla_{\rho}C^{\alpha\beta\mu\rho}-\frac{1}{3q}Q_{\rho}C^{\alpha\beta\mu\rho}=0
\label{spinvacuum}
\end{eqnarray}
in which $R_{\rho\sigma\mu\nu}$ and $\nabla_{\mu}$ are the curvature and the covariant derivative of the purely metric case given by the Riemann curvature and in terms of the Levi-Civita connection as usual; however we see that metric and torsional degrees of freedom are not decoupled within the conformal gravitational field equations.

Now, considering this system of field equations, and taking the first in its decomposition we have that its antisymmetric part is given by
\begin{eqnarray}
&\frac{1}{3q}(\nabla_{\rho}Q_{\sigma}-\nabla_{\sigma}Q_{\rho})C^{\mu\sigma\alpha\rho}=0
\end{eqnarray}
identically, whose solutions in the case in which the spacetime is not conformally flat reduce to $\frac{1}{3q}(\nabla_{\rho}Q_{\sigma}-\nabla_{\sigma}Q_{\rho})=0$ then implying in a connected and simply connected spacetime that the trace vector is the gradient of a certain scalar field given by $Q_{\sigma}=3q\nabla_{\sigma}\varphi$ removable by means of suitable conformal transformations; in fact by choosing $\varphi\equiv-\phi$ it is possible to see that a conformal transformation would map the trace vector into the vanishing one, thus forcing torsion to be equal to zero: it is important to remark that the torsionless theory has not been obtained by setting torsion to zero by hand, but by exploiting conformal transformation. On the other hand however, we cannot perform further conformal transformations as we have already exhausted the only degree of freedom conceded by the conformal invariance of the theory; the conformal field equations (\ref{energyvacuum}-\ref{spinvacuum}) in torsionless conformal gauge respectively reduce to
\begin{eqnarray}
&C^{\theta\sigma\rho\alpha}C_{\theta\sigma\rho}^{\phantom{\theta\sigma\rho}\mu}
-\frac{1}{4}g^{\alpha\mu}C^{\theta\sigma\rho\beta}C_{\theta\sigma\rho\beta}
+C^{\mu\sigma\alpha\rho}R_{\sigma\rho}=0
\label{energytorsionless}\\
&\nabla_{\rho}C^{\alpha\beta\mu\rho}=0
\label{spintorsionless}
\end{eqnarray}
which are a system of two field equations of which one is of the second-order derivative and quadratic and the other is linear and of the third-order derivative, differently from what we had in the purely metric case with only one equation of fourth-order derivative. Finally, field equations (\ref{energytorsionless}-\ref{spintorsionless}) may be written as
\begin{eqnarray}
\nonumber
&R^{\theta\sigma\rho\alpha}R_{\theta\sigma\rho}^{\phantom{\theta\sigma\rho}\mu}
-\frac{1}{4}g^{\alpha\mu}R^{\theta\sigma\rho\beta}R_{\theta\sigma\rho\beta}-\\
&-R^{\mu\beta\alpha\rho}R_{\beta\rho}-R^{\mu\rho}R^{\alpha}_{\phantom{\alpha}\rho}
+\frac{1}{2}g^{\alpha\mu}R^{\rho\beta}R_{\rho\beta}
+\frac{1}{3}R^{\alpha\mu}R-\frac{1}{12}g^{\alpha\mu}R^{2}=0
\label{quadratic}\\
&\nabla_{\mu}(R_{\beta\nu}-\frac{1}{6}Rg_{\beta\nu})
-\nabla_{\nu}(R_{\beta\mu}-\frac{1}{6}Rg_{\beta\mu})=0
\label{differential}
\end{eqnarray}
which are the final system of purely metric field equations in our theory.

Next we notice the fundamental fact that by taking the quadratic field equation given by (\ref{quadratic}) and adding the divergence of the differential field equation given by (\ref{differential}) we get the following couple of field equations
\begin{eqnarray}
\nonumber
&R^{\theta\sigma\rho\alpha}R_{\theta\sigma\rho}^{\phantom{\theta\sigma\rho}\mu}
-\frac{1}{4}g^{\alpha\mu}R^{\theta\sigma\rho\beta}R_{\theta\sigma\rho\beta}-\\
\nonumber
&-R^{\mu\beta\alpha\rho}R_{\beta\rho}-R^{\mu\rho}R^{\alpha}_{\phantom{\alpha}\rho}
+\frac{1}{2}g^{\alpha\mu}R^{\rho\beta}R_{\rho\beta}
+\frac{1}{3}R^{\alpha\mu}R-\frac{1}{12}g^{\alpha\mu}R^{2}+\\
&+\nabla^{2}R^{\alpha\mu}-\frac{1}{6}g^{\alpha\mu}\nabla^{2}R
+\frac{1}{6}\nabla^{\alpha}\nabla^{\mu}R-\nabla_{\rho}\nabla^{\mu}R^{\alpha\rho}=0
\label{quadraticdifferential}\\
&\nabla_{\mu}(R_{\beta\nu}-\frac{1}{6}Rg_{\beta\nu})
-\nabla_{\nu}(R_{\beta\mu}-\frac{1}{6}Rg_{\beta\mu})=0
\label{reallydifferential}
\end{eqnarray}
and in (\ref{quadraticdifferential}) in the last line the last term can equivalently be rewritten by means of $\nabla^{\rho}\nabla_{\mu}R_{\alpha\rho}\equiv \nabla_{\mu}\nabla^{\rho}R_{\alpha\rho}+[\nabla^{\rho},\nabla_{\mu}]R_{\alpha\rho}\equiv 
\frac{1}{2}\nabla_{\mu}\nabla_{\alpha}R-R^{\iota\rho}R_{\alpha\iota\mu\rho}
+R_{\alpha\iota}R_{\mu}^{\phantom{\mu}\iota}$ in a form for which we can eventually simplify the field equations as to write them into a known way while field equation (\ref{reallydifferential}) may be left unmodified so the entire system of field equations is eventually written in the following manner
\begin{eqnarray}
\nonumber
&\nabla^{2}R^{\alpha\mu}-\frac{1}{6}g^{\alpha\mu}\nabla^{2}R
-\frac{1}{3}\nabla^{\alpha}\nabla^{\mu}R+\\
\nonumber
&R^{\theta\sigma\rho\alpha}R_{\theta\sigma\rho}^{\phantom{\theta\sigma\rho}\mu}
-\frac{1}{4}g^{\alpha\mu}R^{\theta\sigma\rho\beta}R_{\theta\sigma\rho\beta}-\\
&-2R^{\mu\rho}R^{\alpha}_{\phantom{\alpha}\rho}
+\frac{1}{2}g^{\alpha\mu}R^{\rho\beta}R_{\rho\beta}
+\frac{1}{3}R^{\alpha\mu}R-\frac{1}{12}g^{\alpha\mu}R^{2}=0
\label{differentialquadratic}\\
&\nabla_{\mu}(R_{\beta\nu}-\frac{1}{6}Rg_{\beta\nu})
-\nabla_{\nu}(R_{\beta\mu}-\frac{1}{6}Rg_{\beta\mu})=0
\label{differentialtrue}
\end{eqnarray}
where field equation (\ref{differentialquadratic}) is precisely the form of the field equations we would have obtained in the purely metric case while the field equation (\ref{differentialtrue}) is the one that comes from torsion with no equivalent in the purely metric case, that is from the present field equations in vacuum it is possible to employ conformal transformations to extract a torsionless limit which gives rise to the purely metric field equations we would have had in the purely metric case but also to additional torsional field equations that even in the torsionless limit do not reduce to be trivial; therefore what we have proven before in general receives now an explicit example in the case of conformal theory in the vacuum of matter fields and in the torsionless gauge: the system of field equations we have obtained by varying with respect to vierbein and spin-connection has eventually been decomposed into a system of field equations for metric and torsion that is the one we would have obtained if we were to vary with respect to metric and torsion in the first place, but both systems of field equations are inequivalent from what we would have obtained by varying with respect to the metric alone in the case torsion is never present because of the presence of an additional field equation coming from torsion that remains non-trivial even in the torsionless limit but which would not be present at all in case torsion were instead neglected from the very beginning. This additional field equation is what makes the difference, that is its presence makes the present theory more restricted than the Weyl purely metric conformal gravity, and it is only when the additional field equations actually happens to become trivial that the two approaches really coincide.

We have thus two possible schemes: one in which torsion is later removed via a conformal transformation and one in which torsion is never present: the former has two field equations that in torsionless conformal gauge reduce to be the one we would have had in the purely metric case plus an additional field equation that does not reduce to be trivial and which could have never been obtained in the purely metric case, and because of this additional field equation the theory in which torsion is set to zero is more restrictive than the theory in which torsion is never present; this is true in general, but there could be special cases in which the torsion field equation does become trivial, so that no further constraint is present and the torsionless limit gives the field equations we would have had if torsion were always neglected. So the metric-torsional gravity in the torsionless limit is more restricted than the purely metric gravity in general, although they may happen to be equivalent in some special circumstances.

In the next sub-sections we study these two cases: first the stationary space isotropic with respect to an origin, showing a counterexample proving that in general the theory in which torsion is present and then set to zero is more restrictive than the theory in which torsion is always neglected; then the case in which the space is isotropic and homogeneous, showing that there are special cases in which the two approaches happen to be equivalent.
\subsection{Isotropic Space}
We shall now look for solutions in the case of isotropic space, that is with spherical symmetry; the case we will take into account is static and therefore time independent: in the frame at rest with respect to the origin of the coordinates given by $(t,r,\theta,\varphi)$ the metric is given in terms of two functions of the radial coordinate $A(r)$ and $B(r)$ taken to be positive defined and it can be written as
\begin{eqnarray}
&g_{tt}=A\ \ \ \ \ \ \ \ g_{rr}=-B\ \ \ \ \ \ \ \ g_{\theta\theta}=-r^{2}\ \ \ \ \ \ \ \ g_{\varphi\varphi}=-r^{2}(\sin{\theta})^{2}
\end{eqnarray}
with all other components equal to zero. The Levi-Civita connection is
\begin{eqnarray}
\nonumber
&\Lambda^{t}_{tr}=\frac{A'}{2A}\ \ \ \ \Lambda^{r}_{tt}=\frac{A'}{2B}\ \ \ \  \Lambda^{r}_{rr}=\frac{B'}{2B}\ \ \ \ \ \ \Lambda^{r}_{\theta\theta}=-\frac{r}{B}\ \ \ \ \ \ \Lambda^{r}_{\varphi\varphi}=-\frac{r}{B}(\sin{\theta})^{2}\\
&\Lambda^{\theta}_{\theta r}=\frac{1}{r}\ \ \ \ \ \ 
\Lambda^{\theta}_{\varphi\varphi}=-\cot{\theta}(\sin{\theta})^{2}\ \ \ \ \ \ 
\Lambda^{\varphi}_{\varphi r}=\frac{1}{r}\ \ \ \ \ \ 
\Lambda^{\varphi}_{\varphi\theta}=\cot{\theta}
\end{eqnarray}
with all other components vanishing and the Riemann curvature is 
\begin{eqnarray}
\nonumber
&R^{t}_{\phantom{t}rtr}=-\frac{A''}{2A}+\frac{A'^{2}}{4A^{2}}+\frac{A'B'}{4AB}\ \ \ \ R^{t}_{\phantom{t}\theta t \theta}=-\frac{A'r}{2AB}\ \ \ \ 
R^{t}_{\phantom{t}\varphi t \varphi}=-\frac{A'r}{2AB}(\sin{\theta})^{2}\\
&R^{r}_{\phantom{r}\theta r \theta}=\frac{B'r}{2B^{2}}\ \ \ \ 
R^{r}_{\phantom{r}\varphi r \varphi}=\frac{B'r}{2B^{2}}(\sin{\theta})^{2}\ \ \ \
R^{\theta}_{\phantom{\theta}\varphi\theta\varphi}=\left(1-\frac{1}{B}\right)(\sin{\theta})^{2}
\end{eqnarray}
with contraction
\begin{eqnarray}
\nonumber
&R^{t}_{\phantom{t}t}
=\frac{A''}{2AB}-\frac{A'^{2}}{4A^{2}B}-\frac{A'B'}{4AB^{2}}+\frac{A'}{ABr}\ \ 
R^{r}_{\phantom{r}r}
=\frac{A''}{2AB}-\frac{A'^{2}}{4A^{2}B}-\frac{A'B'}{4AB^{2}}-\frac{B'}{B^{2}r}\\
&R^{\theta}_{\phantom{\theta}\theta}
=\frac{A'}{2ABr}-\frac{B'}{2B^{2}r}+\frac{1}{Br^{2}}-\frac{1}{r^{2}}\ \ 
R^{\varphi}_{\phantom{\varphi}\varphi}
=\frac{A'}{2ABr}-\frac{B'}{2B^{2}r}+\frac{1}{Br^{2}}-\frac{1}{r^{2}}
\end{eqnarray}
and contraction
\begin{eqnarray}
&R=\frac{A''}{AB}-\frac{A'^{2}}{2A^{2}B}-\frac{A'B'}{2AB^{2}}
+\frac{2A'}{ABr}-\frac{2B'}{B^{2}r}+\frac{2}{Br^{2}}-\frac{2}{r^{2}}
\end{eqnarray}
all other components vanishing identically. From these we can compute all covariant derivatives and squared curvatures in the field equations.

With this symmetry, field equations (\ref{quadratic}) are diagonal and in them the component $(\varphi\varphi)$ is proportional to the component $(\theta\theta)$, so that by employing the tracelessness of these field equations we see that also the component $(\theta\theta)$ can be written in terms of components $(rr)$ and $(tt)$; component $(rr)$ is solved in terms of $(tt)$ once the constraint $AB=1$ is imposed: what now is left as only component $(tt)$ can be written after calling $A=1+h$ in a form that is factorized in two linear field equations of the second-order derivative. With these symmetries and for the assumptions above, also the field equations (\ref{differential}) reduce to have the single independent component given by $(rtt)$ in the form of an integral of the field equations. The general solution can then be found by integrating this field equation (\ref{differential}) and later by constraining the obtained solution employing the constraining equation (\ref{quadratic}) or by solving (\ref{differential}) and (\ref{quadratic}) simultaneously.

However, field equations (\ref{differential}-\ref{quadratic}) are equivalent to (\ref{differentialtrue}-\ref{differentialquadratic}), which are more restricted than the field equations of the standard theory (\ref{differentialquadratic}) alone because of the presence of the additional field equation (\ref{differentialtrue}); therefore the solution we are looking for can also be obtained by considering the solutions of the standard theory and imposing they satisfy also (\ref{differentialtrue}): this strategy is easier because we already know the most general solutions of the standard theory \cite{m-k}
\begin{eqnarray}
&h=-\frac{2m(1-3m\gamma)}{r}-6m\gamma+2\gamma r-kr^{2}
\label{solutionMK}
\end{eqnarray}
so that we only need to insist that this solution is constrained by (\ref{differentialtrue}), after which it is easy to see that the most general solution of the present theory is given by (\ref{solutionMK}) with $m\gamma\!=\!0$ as constraint. We have two alternatives: either 
\begin{eqnarray}
&h=-\frac{2m}{r}-kr^{2}
\label{solution1}
\end{eqnarray}
which is asymptotically Schwarzschild for short distances and asymptotically de Sitter for large distances with constant spacetime curvature $-12k$; or
\begin{eqnarray}
&h=2\gamma r-kr^{2}
\label{solution2}
\end{eqnarray}
which is conformal to the Friedmann-Lema\^{\i}tre-Robertson-Walker solution with constant spatial curvature $-k$ for which the Weyl tensor is zero identically.

We shall now compare the two approaches: we notice that in \cite{m-k} Mannheim and Kazanas have a unique solution able to describe both the Schwarzschild limit for solar system scales and the Friedmann-Lema\^{\i}tre-Robertson-Walker limit for cosmological scales, and in addition as a by-product of the simultaneous presence of both terms it is also capable to fit the rotation curves for intermediate galactic scales; here instead we do not have such intermediate behaviour in either solutions, and the short and large distance behaviours are described independently by two disconnected solutions of which one is only the FLRW solution without the Schwarzschild limit near the origin. However, we do not regard this as a weak point of our approach; to explain why we first remark that any meaningful quantity in conformal relativity is to be constituted by conformal tensors, that is the Weyl curvature: if we calculate the Weyl tensor in these examples we see that for solution (\ref{solutionMK}) it is not zero even at infinity where a constant value proportional to the product $m\gamma$ is reached, whereas for solution (\ref{solution1}) it drops to zero at infinity and solution (\ref{solution2}) it is always zero. If we were to assume the reasonable hypothesis that all relevant quantities of a theory should vanish at infinity then $m\gamma\!=\!0$, which is precisely what we automatically have in the present cases for both solutions (\ref{solution1}) or (\ref{solution2}); and if we neglect the trivial solution then there will be the solution (\ref{solution1}) alone. This solution interpolates the Schwarzschild one near the origin and the de Sitter one for large scales.
\subsection{Isotropic-Homogeneous Space}
We shall next look for solutions in the case of homogeneous and isotropic space with spatial curvature $k$; this case has only temporal dependence: with coordinates $(t,r,\theta,\varphi)$ the metric is given in terms of only one function of the cosmological time $A(t)$ and it can be written according to the following form
\begin{eqnarray}
&g_{tt}=1\ \ \ \ g_{rr}=-\frac{A^{2}}{\left(1-kr^{2}\right)}\ \ \ \ 
g_{\theta\theta}=-A^{2}r^{2}\ \ \ \ g_{\varphi\varphi}=-A^{2}r^{2}(\sin{\theta})^{2}
\end{eqnarray}
with all other components equal to zero. The Levi-Civita connection is
\begin{eqnarray}
\nonumber
&\Lambda^{t}_{rr}=\frac{A\dot{A}}{\left(1-kr^{2}\right)}\ \ \ \ \ \ 
\Lambda^{t}_{\theta\theta}=A\dot{A}r^{2}\ \ \ \ \ \ 
\Lambda^{t}_{\varphi\varphi}=A\dot{A}r^{2}(\sin{\theta})^{2}\\
\nonumber
&\Lambda^{r}_{rt}=\frac{\dot{A}}{A}\ 
\Lambda^{r}_{rr}=\frac{kr}{\left(1-kr^{2}\right)}\ 
\Lambda^{r}_{\theta\theta}=\left(kr^{3}-r\right)\ 
\Lambda^{r}_{\varphi\varphi}=\left(kr^{3}-r\right)(\sin{\theta})^{2}\\
\nonumber
&\Lambda^{\theta}_{\theta t}=\frac{\dot{A}}{A}\ \ \ \ \ \ \ \ 
\Lambda^{\theta}_{\theta r}=\frac{1}{r}\ \ \ \ \ \ \ \ 
\Lambda^{\theta}_{\varphi\varphi}=-\cot{\theta}(\sin{\theta})^{2}\\
&\Lambda^{\varphi}_{\varphi t}=\frac{\dot{A}}{A}\ \ \ \ \ \ 
\Lambda^{\varphi}_{\varphi r}=\frac{1}{r}\ \ \ \ \ \ 
\Lambda^{\varphi}_{\varphi\theta}=\cot{\theta}
\end{eqnarray}
all other components vanishing and the Riemann curvature is
\begin{eqnarray}
\nonumber
&R^{t}_{\phantom{t}rtr}=\frac{A\ddot{A}}{\left(1-kr^{2}\right)}\ \ \ \ 
R^{t}_{\phantom{t}\theta t \theta}=A\ddot{A}r^{2}\ \ \ \ 
R^{t}_{\phantom{t}\varphi t \varphi}=A\ddot{A}r^{2}(\sin{\theta})^{2}\\
\nonumber
&R^{r}_{\phantom{r}\theta r \theta}=\left(\dot{A}^{2}+k\right)r^{2}\ \ \ \ 
R^{r}_{\phantom{r}\varphi r \varphi}=\left(\dot{A}^{2}+k\right)r^{2}(\sin{\theta})^{2}\\
&R^{\theta}_{\phantom{\theta}\varphi\theta\varphi}=\left(\dot{A}^{2}+k\right)r^{2}(\sin{\theta})^{2}
\end{eqnarray}
with contraction
\begin{eqnarray}
\nonumber
&R^{t}_{\phantom{t}t}=-3\left(\frac{\ddot{A}}{A}\right)\ \ \ \ 
R^{r}_{\phantom{r}r}=-2\left(\frac{\ddot{A}}{2A}+\frac{\dot{A}^{2}}{A^{2}}+\frac{k}{A^{2}}\right)\\
&R^{\theta}_{\phantom{\theta}\theta}
=-2\left(\frac{\ddot{A}}{2A}+\frac{\dot{A}^{2}}{A^{2}}+\frac{k}{A^{2}}\right)\ \ \ \ 
R^{\varphi}_{\phantom{\varphi}\varphi}
=-2\left(\frac{\ddot{A}}{2A}+\frac{\dot{A}^{2}}{A^{2}}+\frac{k}{A^{2}}\right)
\end{eqnarray}
and contraction
\begin{eqnarray}
&R=-6\left(\frac{\ddot{A}}{A}+\frac{\dot{A}^{2}}{A^{2}}+\frac{k}{A^{2}}\right)
\end{eqnarray}
all other components vanishing identically. And from these quantities we can calculate all covariant derivatives and square curvatures in the field equations.

In this case, the metric is conformal to the solution (\ref{solution1}) and (\ref{solution2}) with both coefficients $m\!=\!\gamma\!=\!0$, whose conformal tensor of curvature vanishes; so this is always a solution, although trivial. This shows that there are in fact situations in which the two approaches happen to be equivalent.
\section*{Conclusion}
In this paper, we have considered the metric-torsional conformal gravity in vacuum, and we have studied a specific case in which torsion was equivalent to its trace vector, eventually removable via a conformal transformation by choosing what we have called the torsionless gauge, and we have studied two cases of particular symmetries: in the case of isotropy, we found that a solution could be given as a restriction of the solution found in \cite{m-k}, that is as solution of the type (\ref{solutionMK}) with constraint $m=0$ or $\gamma=0$, and so of the type (\ref{solution1}) having a Schwarzschild behaviour for short distances and a de Sitter behaviour for large distances or of an alternative type (\ref{solution2}) which is trivial; in the case of isotropy and homogeneity, we found that the metric is the one we had in the previous case with both $m=0$ and $\gamma=0$, and so like (\ref{solution2}) trivial. Eventually, in the future it would be interesting to know what happens for gravitational waves.

The solutions obtained here are more restrictive than the solutions given in the standard situation, although in some cases it happens that they are equivalent; that is the solutions given in models in which torsion is present and later removed in the torsionless gauge are more constrained than those in which torsion is always zero in general: this is due to the presence of an additional field equation that does not become trivial necessarily. This proves that higher-order derivative gravitational torsional theories do not necessarily possess a torsionless limit that is continuous in the field equations.

\end{document}